\documentclass[twocolumn,10pt]{article}
\usepackage[utf8]{inputenc}
\pdfoutput=1

\usepackage[
  paper=letterpaper,
  lmargin=0.8in,
  rmargin=0.8in,
  tmargin=1in,
  bmargin=1in]{geometry}
  
\usepackage[compact]{titlesec}
\titlespacing{\section}{0pt}{*0}{*0}
\titlespacing{\subsection}{0pt}{*0}{*0}
\titlespacing{\subsubsection}{0pt}{*0}{*0}

\usepackage[pdftex,final]{graphicx}
\usepackage[pdftex]{color}

\usepackage{authblk}

\usepackage{fancyhdr}
\usepackage[colorlinks=true,urlcolor=black,citecolor=black,linkcolor=black,final=true]{hyperref}

\usepackage[font={small,bf},textfont=md,format=hang,margin=10pt,singlelinecheck=false]{caption}

\title{Sustainable Software Ecosystems for Open Science: \\ 15 Years of Practice
and Experience at Kitware}
\author{Marcus D. Hanwell}
\author{Amitha Perera}
\author{Wes Turner}
\author{Patrick O'Leary}
\author{Katie Osterdahl}
\author{Bill Hoffman}
\author{Will Schroeder}
\affil{Kitware, Inc., 28 Corporate Drive, Clifton Park, NY 12065 USA.\\ \url{http://www.kitware.com/}}

\begin{document}

\maketitle

\section*{Introduction}

Mathematics is the core language of science, and for centuries it was necessary 
to show the mathematical underpinnings of new research as part of scientific
explorations. This lingua franca provided an essential level of
understandability and precision; providing for unambiguous communication and
rigorous verification of scientific claims beyond the inaccuracies of spoken 
languages. However, the last few decades have seen an erosion in this paradigm.
The increased reliance of science on complex computational codes and large data
makes the description of all but the most basic research error prone,
impenetrable, and unverifiable. 

This issue is not restricted to just one field of science, but is endemic
throughout the broader scientific community and the consequences of opaque 
processes and lack of reproducibility are not trivial. Cases of irreproducible
studies and clinical trials have been making headlines, from Bayer Health Care 
stopping nearly two-thirds of its target-validation projects because of 
inconsistencies with the initially published claims, to global economic policy
being based on a single fundamentally flawed study by Harvard economists. These 
costly mistakes can be remedied much earlier, and before key decisions are made,
simply by returning transparency and precision to the process of publication and
review.

The question we must address is how best to reinstate a common language
and what that language should be. We believe that the only practical choice is
to require that disclosures of scientific research based on complex codes and
data use the very same complex codes and data as the common language of
publication. This means that as new studies and new scientific explorations are
undertaken, the data, methods, and software used by the researchers to arrive
at their conclusions must be made available and accessible to other researchers
and the general populace. If this goal is to be realized then the standard of
software engineering in science must be improved, and sustainable software
ecosystems with meaningful credit must be realized. This is not simply limited
to teaching scientists to write code; if sustainable software projects are to be
established in science then issues such as testing, licensing, and collaboration
must be addressed. Sufficient engineering discipline is required to realize
robust foundations that can be extended through the use of code review,
regression testing and proper citation and attribution of software used in
research.

\section*{Vision}

\href{http://www.kitware.com/}{Kitware} has been developing open-source solutions and laying the building
blocks for open science for over a decade. We contribute to a wide and varied range of
open-source projects in the scientific domain across many disciplines including 
supercomputing, scientific
data visualization, large data management and informatics, computational fluid
dynamics, medical imaging, chemistry and computer vision. Kitware's specific
expertise in this area, including business practices, collaboration patterns,
and growth strategies, are discussed. Based on this experience, we believe
practical frameworks for science revolve around well-defined and open
application programming interfaces (APIs) between key steps in the analysis
workflow from raw data to final figures in published manuscripts.

Although not directly related to the scientific goals, it is important to
consider licensing in the context of scientific communities. It is not enough to
simply make code and data available for free if we wish to see communities
flourish---shared ownership and access must be provided under permissive
licenses wherever possible. These should encourage reuse, derivatives, and allow
those products to be shared. Permissive licenses such as MIT, BSD, and Apache 2
for software; and CC0 or CC-BY for data should be strongly encouraged over
copyleft, non-commercial and no derivative clauses that can hamper further reuse
of code or data. When a wealth of data and code is available under permissive
licenses that encourage reuse it is much easier to grow sustainable software
ecosystems---suddenly the method a student developed in their research can be
incorporated into a larger framework even after a student has moved on to the
next challenge. This relies upon the code being published and made available
under permissive licenses allowing for reuse in other contexts. The same path
can be considered for data, where individual findings can be combined into
larger collections and analyses performed on aggregate data to make new
discoveries outside the scope of the original research in many cases.

\section*{Practice and Experience}

Kitware started as an experiment in open source education beginning when three
General Electric research employees wanted to author a book on visualization using a new
programming language called C++. The book included code, and in order to 
encourage reuse and improve the general standard in the field they were granted
authorization by GE to retain copyright and publish the code with the book. This would
later become the Visualization Toolkit (VTK). In 1998 two of the authors left GE
and with several others founded a small company in upstate New York named
Kitware. Fifteen years later the adventure continues, with over 100 employees
and sustained year-on-year growth.

\subsection*{Baseline Implementations} 

As noted, VTK began life as code to support a book. The goal was to improve
the state-of-the-art in the visualization community by putting tested and
verified implementations of important visualization and analysis algorithms into
a software library that was permissively licensed and available for use by a
broad and open community. The intent was to provide these implementations as a
concrete instantiation of complex algorithms so that researchers around the 
world did not have to do this from the papers of the time, with varying degrees
of success and different levels of verification. This same goal lies at the
heart of the Insight Toolkit in medical imaging for image segmentation and 
registration. Over the years things such as the Insight Journal were added so 
that new algorithms could be proposed in formal publications that included code,
data, and baselines demonstrating the algorithm. If accepted, the algorithm would
be merged into the main codebase and made available to the wider community.

\subsection*{Business Model}

The majority of well-known software companies use a licensing and intellectual
property model in order to derive revenue from their development activity. In
the sciences this is often bolstered by funding from major agencies as well. Kitware
generates no intellectual property, and has moved away from the licensing model
to a large extent---focusing our business activity on a services model. This is
an extremely effective strategy when working in high-performance computing, and
addresses one of the major flaws in many open-source projects where support is
difficult to find. Services make up a larger share of the industry sector, with
companies such as IBM and Red Hat deriving most of their income from
services rather than licensing. This model enables us to work with academics,
national laboratories, and industry as partners often with joint funding streams
to develop new features and make headway in different avenues of research and
development. Through the use of permissive licensing models, rigorous software
processes and agile development methodologies we have been able to make
significant progress and received recognition for major projects such as VTK
and ParaView.

\subsection*{Developing Communities} 

\begin{figure}[!hbtp]
\centering
\includegraphics[width=0.45\textwidth]{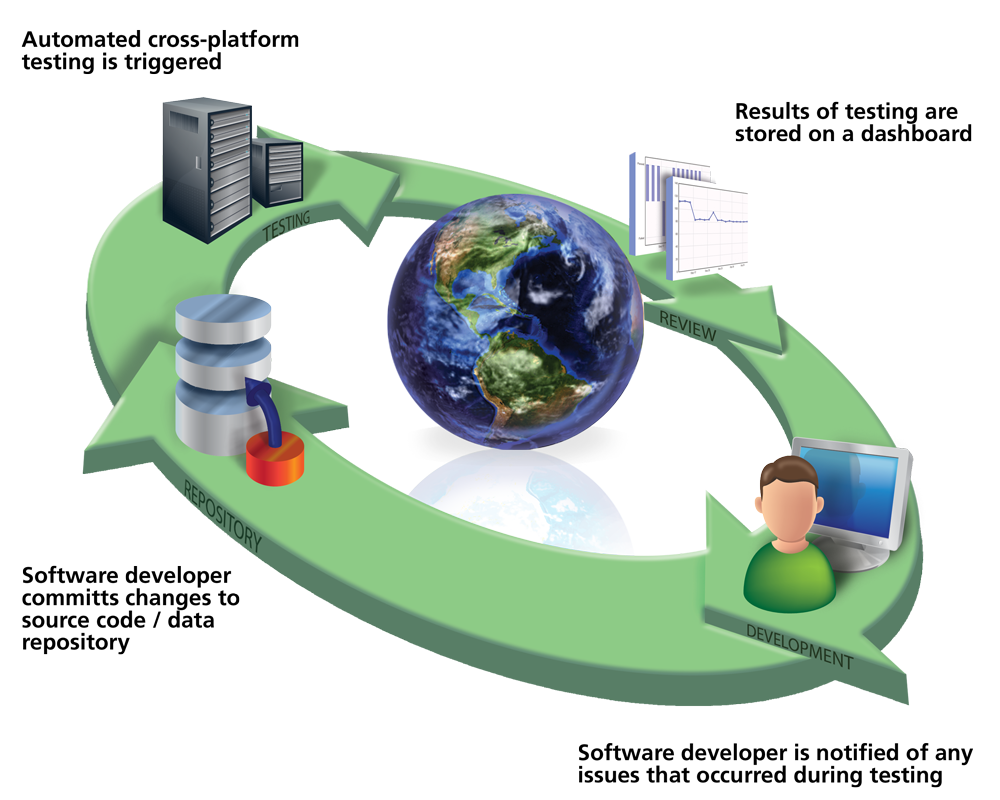}
\caption{Overview of the Kitware software process.}
\label{fig:software-process}
\end{figure}

One of our core focus areas at Kitware is software process and developing robust
communities around software projects that can grow beyond individual contracts.
This process is shown at a high level in Figure~\ref{fig:software-process} where
developers commit code, it is automatically tested, dashboards gather testing
information and disseminate that information.
This process is powered by CMake, which began as a build system for ITK when existing build systems 
proved inadequate, with the simple goal of building C++ code on all major 
platforms. This was later augmented with CTest that would run and report the 
results of automated testing to Dart, and later CDash. Packaging was addressed
using CPack, enabling some degree of abstraction when creating binary installers
which are now routinely generated each night so that non-developers can
install the latest build of a software project to see if a bug has been fixed.

Kitware has built up software workflows around version
control systems such as RCS, CVS, Subversion and more recently Git (among other
distributed version control systems). These processes are designed to minimize
the barrier to entry while maintaining the high quality of the software using
automated repository hooks, continuous integration, and advanced testing
capabilities such as image-based regression images to verify that visualization
algorithms produce consistent and correct output on a large number of platforms
beyond the capability of any individual contributor.

Any community needs to establish a strong process to remain viable over the long
term. This requires a mixture of technical and social resources to facilitate
productive engagement and growth. Scientific projects often have limited
resources, with a high turnover of developers/engineers as they move through
their academic careers. This means that it is essential to establish procedures
to help new community members get up to speed, and to retain group knowledge as 
people move on to new positions. Some will maintain long-term ties through
different positions, as is normal in more mainstream open-source projects, 
others will not and it is important to ensure the parts of the project they 
developed do not atrophy.

\subsection*{Open Software Process}

\begin{figure}[!hbtp]
\centering
\includegraphics[width=0.45\textwidth]{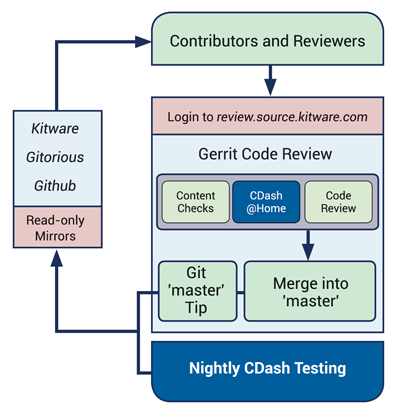}
\caption{VTK's software process incorporating online code review, automated
testing of topic branches, and nightly testing with distributed mirrors.}
\label{fig:software-process2}
\end{figure}

Kitware engineers and researchers have experimented with a number of techniques
to improve induction of new community members and employees into our company and
software communities, to support open-source development, and the growth of
our projects. Most importantly, these experiments are more than just an
academic exercise. All of the techniques are applied as part of our ongoing,
commercial, software business. The techniques that offer pragmatic benefits to
our business and open-source communities survive. Figure~\ref{fig:software-process2}
shows a more detailed view of the software process used by the Visualization
Toolkit (VTK).

Recently we started using Gerrit to conduct online code reviews, and
added automated triggers so that proposed changes are built on all major 
operating systems. This simple modification allows us to highlight potential
platform-specific issues before engineers review code, saving scarce 
engineering time. Our software projects are developed using test-driven 
development methodologies, and require the submission of tests for new
code or features ensuring that these features are automatically tested and
verified before they are merged.

We have a strong commitment to testing. We run nightly dashboards to provide
wider coverage, ensuring that any problems are normally highlighted within 24 
hours of merging them into the repository and allowing developers to review and 
remedy regressions while the changes are still fresh in their mind. All of these
processes are learning opportunities, enabling developers to quickly discover
what will and won't work with different environments as new features are
developed. These advantages were only made possible in recent years by migrating
to distributed version control systems that enable development outside of the
main repository and for changes to be pushed to staging locations before being
merged into the main development branches.

We have worked hard to close the gap between developers of the frameworks
and applications and their users by automatically generating binary installers.
This enables them to download the latest nightly binary of any given
project to check if their bug still exists, or work more closely with developers
on exercising new features before the final release is made. Underlying all of
this are multiple communication mediums including mailing lists, wikis, and bug
trackers to offer a comprehensive set of community tools. Having a commercial
company backing up these projects also offers the ability to provide books,
webinars, and on-site courses.

\subsection*{Mature Communities}

\begin{figure}[btp]
\centering
\includegraphics[width=0.4\textwidth]{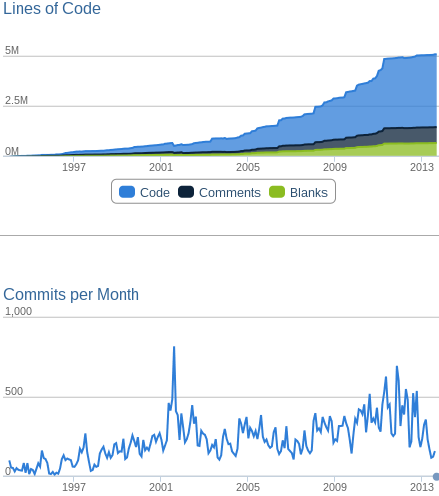}
\caption{Statistics generated by Ohloh showin lines of
code, and commit rates for VTK.}
\label{fig:ohloh}
\end{figure}

Kitware hosts or plays a leading role in the maintenance and engineering on a
number of projects, and over the years has learned to be adaptable. The oldest
Kitware-led community is VTK, which also has one of the longest running version 
histories of any project whose history is publicly available. This is a
community that grew up organically around the VTK book, and many pieces of the 
code were experimented with and added as the community grew. According to Ohloh 
the project has had nearly 55,000 commits made by over 200 contributors and
includes more than 3.5 million lines of code. The first recorded VTK commit 
occurred nearly 20 years ago and development remains active. Even when looking
at the last 12 months the project has seen over 3,000 commits from 66 
contributors under a permissive BSD license. Figure~\ref{fig:ohloh} shows a
graphical overview of the project's activity from 1994 to present.

Even so, these numbers understate the total significance. For instance, they do
not count the number of commercial contracts where Kitware supported the toolkit
by providing engineers to extend VTK, or create/augment focused applications 
that leverage VTK, or research conducted using VTK.
The development is also funded by several ongoing contracts 
with established groups, and funding streams from various federal agencies such 
as the DOE, NIH, DOD, DARPA, and NSF. Due to VTK's importance to researchers in
the medical area, it was granted a rare R01 maintenance grant from the NIH.
Under this contract a consortium of
companies and universities led by Kitware is working to update and overhaul 
aspects of the VTK rendering subsystems and to make other improvements over the
next four years. VTK was able to garner a large number of support letters from 
the community of users and developers making use of VTK in their research, 
applications and products.

After the success of VTK, the ITK project was, at least in part, founded to 
emulate the success of VTK in the medical segmentation and registration field.
ITK receives ongoing funding from the National Library of Medicine, initially to
provide algorithms that could operate on and make sense of the data from the 
visible human project. It is a little younger, with its first commit in 2000, 
and is interesting as it was one of the first major projects that began after 
Kitware's founding and was designed from the ground up as an open-source project. It 
has seen nearly 45,000 commits in that time from over 200 contributors with 
over 1.5 million lines of code. Over the last year more than 1300 commits were made 
by 67 contributors under a permissive Apache 2 license. The ITK project also
has an associated non-profit organization that is used to coordinate community 
efforts and manage some of the funding and development.

The history of some of these older projects that have seen one or
two decades of development effort should be contrasted with younger projects at
different stages of development.
Experimentation is encouraged, and not all of the projects that are started will
achieve the critical mass required to become successful open source projects, 
but our established processes seek to make experimentation easy and the 
permissive open-source licensing means that even ``failed'' projects will often
live on in other forms if the code developed for the project is useful.

A brief summary of some of these projects include Arbor/Avatol---an NSF funded
project where Kitware is an engineering subcontractor in a larger project led 
by the University of Idaho. The goals of this project are to provide tools for
the analysis and exploration of the Tree of Life. The XDATA project is a
new project funded by DARPA with a large consortium (led by Kitware
as the prime contractor) tasked with creating open-source technology to address
big data analytics. The Open Chemistry project was funded by a DOD SBIR awarded
to Kitware, with the
aim of creating a suite of desktop applications for computational chemists.

In addition to creating, growing and shepherding our own communities Kitware has
more recently begun helping other organizations to build successful communities.
This includes the VA with the OSEHRA project, NA-MIC, and caBIG, all of whom 
use Kitware's expertise to assist them in growing vibrant and successful communities 
around their projects.

\subsection*{Concluding Remarks}

Sustainable software ecosystems are difficult to build, and require concerted
effort, community norms and collaborations. In science it is especially important
to establish communities in which faculty, staff, students and open-source
professionals work together and treat software as a first-class product of
scientific investigation---just as mathematics is treated in the physical sciences.
Kitware has a rich history of establishing collaborative projects in the science,
engineering and medical research fields, and continues to work on improving that
model as new technologies and approaches become available. This approach closely
follows and is enhanced by the movement towards practicing open, reproducible
research in the sciences where data, source code, methodology and approach are all
available so that complex experiments can be independently reproduced and verified.

\subsection*{License}

This document is released under the Creative Commons Attribution 3.0 license (CC-BY),
see \url{http://creativecommons.org/licenses/by/3.0/}.

\end{document}